\documentclass[12pt]{article}
\usepackage{a4}
\usepackage{amsmath}
\usepackage{amssymb}
\usepackage{graphicx}
\usepackage{dcolumn}
\usepackage{bm}
\usepackage{citesort}
\begin{document}

\title{How do trehalose, maltose and sucrose influence some\\
structural and dynamical properties of lysozyme~?\\
An insight from Molecular Dynamics simulations}

\author{A. Lerbret(1), P. Bordat(2), F. Affouard (3), A. H\'edoux (3), Y. Guinet (3),\\
M. Descamps (3)\\ 
(1) Department of Food Science, Cornell University, \\
Stocking Hall, Ithaca, New York 14853 \\
(2) Institut Pluridisciplinaire de Recherche sur l'Environnement et les Mat\'eriaux, \\
UMR 5254, Universit\'e de Pau et des Pays de l'Adour, \\
2 Avenue Pierre Angot, 64053 Pau Cedex 9, France \\
(3) Laboratoire de Dynamique et Structure des Mat\'eriaux Mol\'eculaires, \\
UMR CNRS 8024, Universit\'e Lille I, 59655 Villeneuve d'Ascq Cedex, France}
\date{}
\maketitle

\newpage

\begin{abstract}
The influence of three well-known disaccharides, namely trehalose, maltose and sucrose, on some structural 
and dynamical properties of lysozyme has been investigated by means of 
molecular dynamics computer simulations in the 37-60 wt~\% concentration range.
The effects of sugars on the protein conformation are found 
relatively weak, in agreement with the preferential hydration of lysozyme. Conversely, sugars seem to increase 
significantly the relaxation times of the protein. These 
effects are shown to be correlated to the fractional solvent accessibilities of lysozyme residues and further support 
the slaving of protein dynamics. Moreover, a significant increase in the relaxation times of lysozyme, sugars and 
water molecules is observed within the studied concentration range and may result from the percolation of the 
hydrogen-bond network of sugar molecules. This percolation appears to be of primary importance to explain the influence 
of sugars on the dynamical properties of lysozyme and water. 
\end{abstract}


\newpage

\section{Introduction}

\begin{figure}[!h]
\includegraphics[width=7.2cm,clip=true]{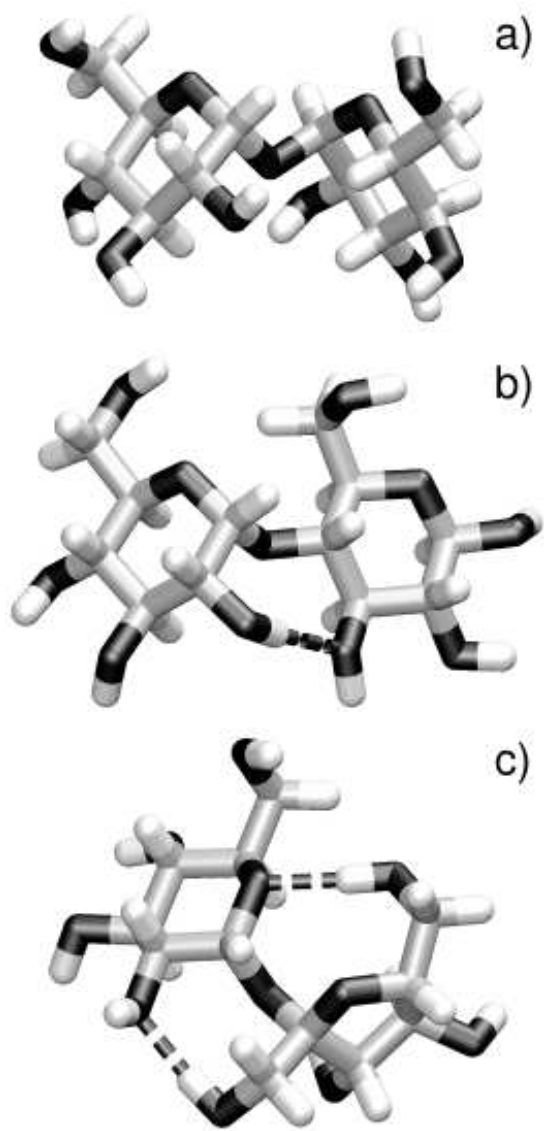}
\caption{\label{sugars}
Schematic representation of trehalose (a), $\beta$-maltose (b) and sucrose (c) in their most stable
cristalline forms, which are dihydrate~\cite{Taga1972}, monohydrate~\cite{Gress1977} and anhydrous~\cite{Brown1973} 
crystals, respectively. Oxygen, carbon and hydrogen atoms are displayed in black, grey and white, respectively.
Intramolecular hydrogen bonds (HBs) are represented by dashed cylinders. In these conformations, trehalose, 
maltose and sucrose form zero, one and two intramolecular HBs, respectively. This figure was generated using
VMD~\cite{Schulten1996} (http://www.ks.uiuc.edu/Research/vmd/).}
\end{figure}

The preservation of biological molecules like proteins is a fundamental goal of pharmaceutical, food, and cosmetic 
industries~\cite{Franks1985,Crowe2001,Wang1999,Haydon2000}. Indeed, they generally remain stable only under very 
stringent conditions of temperature, pH, hydration level or ionic strength~\cite{Wang1999,Wang2000}. The addition of 
stabilizing agents is a common strategy to increase the stability of proteins both in solution or in the dry 
state~\cite{Wang1999,Wang2000}. Among these compatible solutes, trehalose (see figure~\ref{sugars}a), 
a disaccharide ($C_{12}H_{22}O_{11}$), 
has been found particularly efficient~\cite{Crowe1998,Crowe2001}. This sugar is naturally synthesized in large amounts 
by plants, insects and microorganisms such as yeasts and nematodes which are able to withstand for an extended period 
severe environmental conditions of low/high temperatures and/or drought experienced in desert or polar 
regions~\cite{Patist2005,Watanabe2006}. These organisms enter into a biological state where their activity is almost 
completely suspended as long as environmental conditions remain deleterious, and then resume their normal activity. 
This phenomenon is called \textit{anhydrobiosis}. 
Despite the many experimental~\cite{Carpenter1989,Petrelski1993b,Allison1999,Crowe1998} 
and numerical~\cite{Lins2004,Cottone2005,Dirama2006} studies which have been devoted to the \textit{bioprotection} 
phenomenon, the molecular mechanisms at its origin are still not clearly understood. Several hypotheses, mostly based 
on the properties of trehalose - high solubility, low reactivity, good glass-forming and antioxydant properties, 
etc. - have been 
proposed to explain why it is the most effective bioprotectant among sugars and polyols~\cite{Watanabe2006}. 
Nonetheless, none of them is fully satisfactory, since it generally covers only a limited range of temperatures and 
hydration levels. 

In aqueous solutions proteins may be thermodynamically stabilized by the \textit{preferential exclusion} of cosolvent 
molecules from the protein/solvent interface, which make thermodynamically unfavorable the increase of the solvent 
accessible surface area (SASA) of proteins upon denaturation~\cite{Timasheff1995}. This exclusion could arise from 
excluded volume effects and an increase in the surface tension of water induced by solutes~\cite{Bhat2003}. 
This may explain why among 
many solutes trehalose was found to be the most excluded~\cite{Xie1997,Timasheff2002}, given its larger hydration 
number in comparison to other disaccharides such as maltose and sucrose~\cite{Branca2001,Furuki2002} 
(the structures of these sugars are shown in figures~\ref{sugars}b and c, respectively). 
Moreover, solutes may hinder 
the formation of ice, which is generally lethal to organisms. Magaz\`u \textit{et al.} have suggested that trehalose 
would be more effective in inhibiting ice formation than other disaccharides, because it binds to a larger number of 
water molecules and would therefore have a greater \textit{destructuring effect}~\cite{Branca1999b}.

At high temperatures or high osmotic pressures, dehydration is observed. Biological molecules may experience important 
stresses when \textit{hydration} water molecules are removed~\cite{Hoekstra2001}. The effect of the addition of 
stabilizing solutes may be two-fold. First, the solution is likely to vitrify before the complete removal of hydration 
water molecules, since the solution viscosity raises exponentially when decreasing the water content. The glass formed 
may kinetically maintain the conformation of proteins and prevent the fusion of membranes. Trehalose would then be a 
better bioprotectant than maltose or sucrose given its higher glass transition temperature $T_{g}$, as suggested by 
Green and Angell~\cite{Green1989}. This hypothesis is generally well accepted, even though counter-examples 
have been reported~\cite{Allison1999,Cicerone2004,Caliskan2004}.
The second effect of an increase of the concentration of stabilizing solutes upon dehydration is the formation of
hydrogen bonds (HBs) between solutes and biological molecules. Crowe \textit{et al.} have proposed that solutes were 
able to substitute to the hydration water of biological molecules, therefore stabilizing their solution structure and 
functionality~\cite{Crowe1998}. 
This has been supported by the results from many studies~\cite{Carpenter1989,Petrelski1993b,Allison1999}. There is a 
lot of debate on the relative importance of the solvent vitrification and the formation of solute-biomolecule HBs. 
Nevertheless, both appear necessary to achieve long-term preservation~\cite{Crowe1998,Wang2000}.

Alternatively, Belton and Gil have suggested that the direct interaction of the solute glassy matrix with lysozyme is
relatively limited, because the hydration water is trapped in an amorphous state (non-crystalline ice) by the solute 
glassy matrix~\cite{Belton1994}. The role of 
trehalose would be to concentrate residual water molecules close to the protein. This would be consistent with the 
hypothesis of the preferential hydration of proteins in dilute to semi-dilute solutions~\cite{Timasheff2002}.

Finally, Ces\`aro~\textit{et al.}~\cite{Cesaro2001} have pointed out the possible role of the interconversion 
between the trehalose dihydrate crystal $T_{2H_{2}O}$~\cite{Taga1972} and the metastable $\alpha$ phase 
$T_{\alpha}$~\cite{Sussich1998} upon dehydration. This interconversion may make easier the rehydration of biomolecules 
and occurs on time scales compatible with the anhydrobiotic protection, and thus may not induce fast changes in the 
volume or internal pressure of cells, in contrast to ice formation.

Many simulation studies of ternary systems have been published in the last few 
years~\cite{Sum2003,Lins2004,Lins2004,Cottone2005,Doxastakis2005,Skibinsky2005,Dirama2006,Pereira2006}.
Studies of membrane/sugar/water systems~\cite{Sum2003,Lins2004,Doxastakis2005,Skibinsky2005,Pereira2006} 
indicate that sugars, and trehalose in particular, interact directly with membranes, and may stabilize them by forming 
multiple HBs between several adjacent phospholipids. Furthermore, Lins~\textit{et al.}~\cite{Lins2004} have studied 
trehalose-lysozyme interactions at a concentration of about 18 wt \% and have shown that trehalose molecules do not 
expel hydration water. They have suggested that the presence of lysozyme induces an aggregation of trehalose molecules, 
which trap a thin layer at the protein surface, in agreement with the suggestion of Belton and 
Gil~\cite{Belton1994} for solid lysozyme/trehalose/water samples. Similarly, 
Cottone~\textit{et al.}~\cite{Cottone2002} have shown that trehalose is preferentially excluded from the protein 
surface at 50 and 89 wt \%. They also observed an analogous behavior for sucrose at 89 wt \%, which was found more 
excluded than trehalose~\cite{Cottone2005}. Furthermore, Dirama~\textit{et al.}~\cite{Dirama2006} have studied lysozyme 
in a trehalose glass at different temperatures and shown a strong coupling between lysozyme and trehalose molecules 
correlated with the dynamics of protein-sugar intermolecular HBs.

This paper aims at understanding how globular proteins in their flexible native state~\cite{Gregory1997} are influenced at 
room temperature (300 K) by stabilizing solutes at 
intermediate concentrations, \textit{i.e.} high enough so that the solvent structure and dynamics are quite different 
from those of pure water, but low enough to allow relatively large-scale motions of protein surface atoms. This 
concentration range is particularly important since it may link the protein properties in dilute solutions to those in 
solid matrices. For this purpose, we have probed the influence on some structural and dynamical properties 
of hen egg-white lysozyme, a model globular protein, of three well-known homologous disaccharides, namely trehalose 
[$\alpha$-D-glucopyranosyl-$\alpha$-D-glucopyranoside], 
maltose [4-O-($\alpha$-D-glucopyranosyl)-$\beta$-D-glucopyranoside] and 
sucrose [$\beta$-D-fructofuranosyl-$\alpha$-D-glucopyranoside] (see figure~\ref{sugars}), 
in the 37-60 wt~\% concentration range by means of 
molecular dynamics computer simulations. This specific concentration range was chosen because it corresponds to the 
range of concentrations where the HB network of the present sugars was shown to percolate in our previous numerical 
study of disaccharide/water solutions~\cite{Lerbret2005_2}.

\section{Simulation details}

Molecular Dynamics simulations of hen egg-white lysozyme (hereafter referred to as lysozyme) in sugar/water solutions 
have been performed using the CHARMM program~\cite{Brooks1983}, version 29b1. The all-atom CHARMM22 force 
field~\cite{Mackerell1998} has been used to model the protein. The CSFF carbohydrate force field~\cite{Kuttel2002} has 
been considered for disaccharides and water molecules were represented by the SPC/E model~\cite{Berendsen1987}. The 
production simulations were performed in the isochoric-isothermal (N,V,T) ensemble. The length of all covalent bonds 
involving an hydrogen atom as well as the geometry of water molecules were kept fixed using the SHAKE 
algorithm~\cite{Ryckaert1977}, with a relative tolerance of 10$^{-5}$. A 2-fs timestep has been used to integrate the 
equations of motion with the verlet leapfrog algorithm~\cite{Hockney1970}. 

During the different stages of the simulations, the temperatures have been maintained constant with weak coupling to a 
heat bath (Berendsen thermostat~\cite{Berendsen1984}) with a relaxation time of 0.2 ps. A cutoff radius of 10 \AA \ has 
been used to account for van der Waals interactions, which were switched to zero between 8 and 10 \AA \ . A 
Lennard-Jones potential has been employed to represent van der Waals interactions and Lorentz-Berthelot mixing-rules 
have been used for cross-interaction terms. Electrostatic interactions have been handled by the particle mesh Ewald 
(PME)~\cite{Essmann1995} method with $\kappa$ = 0.32 \AA \ $^{-1}$ and the fast-Fourier grid densities set to 
$\sim$ 1/\AA \ (48 and 64 grid points in the X/Y and Z directions, respectively).

The starting structure of lysozyme was obtained from the X-ray crystal structure solved at 1.33 \AA \ (193L entry of 
the Brookhaven Protein Data Bank)~\cite{Vaney1996}. Most probable charge states at pH 7 were chosen for ionizable 
residues. The total charge of lysozyme (+8 $e$) was then neutralized by uniformly rescaling the charge of protein 
atoms, similarly to ref.~\cite{Sterpone2001}. The disaccharide initial conformations have been deduced from neutron 
and X-ray studies of trehalose~\cite{Taga1972}, maltose~\cite{Gress1977} and sucrose~\cite{Brown1973}. The sugar 
concentrations on a protein-free basis are 37, 50 and 60 wt \%. These concentrations have been purposefully chosen 
based on our previous study of sugar/water solutions~\cite{Bordat2004,Lerbret2005}. Indeed, we showed that the 
relative effect of sugars on water may be distinguished above a threshold concentration of about 40 wt\%. Therefore, 
their relative influence on lysozyme at ambient temperature may be characterized above this concentration. Lysozyme 
and its 142 crystallographic hydration water molecules were first placed in an orthorhombic box with cell parameters 
a = b = 46.7 \AA \ and c = 62.2 {\AA}. Then, disaccharide molecules were located and oriented randomly around lysozyme, 
with minimum sugar-protein and sugar-sugar distance criteria, which ensure an isotropic distribution of sugars around 
lysozyme. Finally, water molecules non-overlapping with either lysozyme or sugars were randomly added in the simulation 
box. Initial configurations were minimized in three steps, keeping first lysozyme and sugars fixed, then keeping only 
lysozyme fixed and finally keeping free all molecules. This minimized configuration was heated to 473 K in the canonic 
ensemble during 1 to 3 ns, while maintaining fixed the conformation of lysozyme to prevent conformational changes. 
This aimed at equilibrating solvent configurations, particularly the position and orientation of sugars. Then, the 
resulting configurations were thermalized at 300K and simulated in the isobaric-isothermal (N,P,T) ensemble. The 
stabilized volume of the simulation box during this simulation was considered to compute the averaged density of the 
system and used to perform the subsequent simulations in the (N,V,T) ensemble. A steepest-descent minimization 
procedure of 1000 iterations was then performed, whilst applying a decreasing harmonic potential on atoms of lysozyme. 
After the minimization procedure, the temperature was raised from 0 to 300 K, with a 5-K temperature increase every 
250 steps. Then, an equilibration at 300 K was performed during about 80 ps. Finally, simulations of 10, 12 and 17 ns 
were performed for the sytems at concentrations of 37, 50 and 60 wt \%, respectively, and configurations were saved 
every 0.25 ps. A control simulation of lysozyme in pure water was done in an analogous way as the one described above. 
In this simulation, the orthorhombic box was directly filled with water molecules. Moreover, this system was not heated 
at 473 K, since water molecules are much more mobile than sugars. The first two and four ns were not considered to 
compute the structural and dynamical properties presented in this paper for the 0-50 and 60 wt \% systems, respectively. 
Table~\ref{table1} summarizes some simulation data for the different systems considered in the present study.

\begin{table*}[htbp]
\centering
\caption{\label {table1}
System compositions (where N$_{L}$, N$_{S}$ and N$_{W}$ denote the number of lysozyme, sugar and water molecules, 
respectively), densities, and equilibration/simulation times for the different sugar concentrations $\phi$ (on a 
protein-free basis). Data corresponding to $\phi$ = 0~wt~\% result from only one simulation of the lysozyme/pure water 
solution.
}
\vspace* {1.0cm}
\begin{tabular}{cccccccc}
\cline {1-5}
\hline
$\phi$ (wt \%) & N$_{L}$/N$_{S}$/N$_{W}$ & \multicolumn{3}{c}{density ($g.cm^{-3}$)} & \multicolumn{3}{c}{Eq./Sim. time (ns) }\\
\cline {3-8}
  &  &\multicolumn{1}{c}{T}&\multicolumn{1}{c}{M}&\multicolumn{1}{c}{S}&\multicolumn{3}{c}{T, M, S}\\
\cline {1-8}
0 & 1/0/3800 & \multicolumn{1}{c}{1.04} & \multicolumn{1}{c}{1.04} & \multicolumn{1}{c}{1.04} & \multicolumn{3}{c}{2/8} \\
37 & 1/85/2800 & \multicolumn{1}{c}{1.16} &\multicolumn{1}{c}{1.16} &\multicolumn{1}{c}{1.15}& \multicolumn{3}{c}{2/8} \\
50 & 1/125/2400 & \multicolumn{1}{c}{1.20} &\multicolumn{1}{c}{1.21} &\multicolumn{1}{c}{1.20}& \multicolumn{3}{c}{2/10} \\
60 & 1/165/2100 & \multicolumn{1}{c}{1.24} &\multicolumn{1}{c}{1.25} &\multicolumn{1}{c}{1.24} & \multicolumn{3}{c}{4/13} \\
\hline
\end{tabular}
\end{table*}

\section{Structural properties.}

\subsection{Lysozyme}
\subsubsection{Protein conformation}

The influence of sugars on the conformation of lysozyme has first been checked by computing the root-mean-square 
deviation (RMSD) from the crystallographic structure~\cite{Vaney1996}, of either the C$_{\alpha}$ carbon atoms of the protein backbone 
or of all protein atoms. The averaged values for all studied systems are shown in table~\ref{control}. The values for 
the pure water system are in fair agreement with those reported in previous studies~\cite{Sterpone2001,Lins2004}, 
which employed the TIP3P water model~\cite{Jorgensen1983}. The presence of sugars at the studied concentrations reduces 
conformational changes of lysozyme, as seen from the significant decrease of the $C_{\alpha}$ and the all-atom RMSDs, 
which tends to be more important when disaccharide concentration increases. A comparable effect has been observed by 
Lins \textit{et al.}~\cite{Lins2004} for lysozyme in presence of trehalose at a concentration of $\sim$ 18 wt \% 
compared to the pure water system, where the all-atom RMSD of lysozyme in presence of trehalose was found about 0.2 
\AA \ lower. This reduction of the RMSDs probably arise from the slowing-down of the 
solvent dynamics observed at these concentrations (see section \ref{dynamics}). No major difference is 
observed among the three sugars, whatever the concentration considered. This suggests that they have a 
similar influence on lysozyme from a structural point of view. This is expected in the framework of the preferential 
hydration hypothesis~\cite{Timasheff2002}, which proposes that sugars are preferentially excluded from the protein 
surface. The conformation of lysozyme in the different solutions has also been characterized by means of its radius 
of gyration 
$R_{g}$ and its total solvent accessible surface area (SASA) (calculated with the program DSSP~\cite{Kabsch1983}). These
parameters are given in the remaining part of table~\ref{control}. 
The $R_{g}$ and SASA of lysozyme 
seem to be slightly larger in the presence of sugars. This may stem from the non-negligible interaction of surface 
residues of lysozyme with sugars, as will be shown in the section~\ref{HB_crit}. Again, a clear distinction among the 
studied sugars does not emerge. These three parameters show that the structure of lysozyme in presence of sugars 
remains relatively close to that in pure water.

\begin{table*}[htbp]
\centering
\caption{\label {control}
Parameters describing the structure of lysozyme~: (i) Root-mean-square deviation (RMSD) from the crystallographic 
structure of C$_{\alpha}$ carbon atoms or of all atoms, (ii) radius of gyration $R_{g}$ of lysozyme, and (iii) total 
solvent accessible surface area (SASA). Standard deviations from mean values are given in parentheses.
}
\vspace* {1.0cm}
\begin{tabular}{cccccc}
\cline {1-5}
\hline
$\phi$ (wt \%) &  & \multicolumn{2}{c}{RMSD (\AA)}  & R$_g$ (\AA) & SASA (\AA$^2$) \\
\cline {3-4}
  & Sugar & Backbone & All atoms &  & \\
\cline {1-6}
0 & & 1.40 (0.13) & 2.27 (0.11) & 14.07 (0.07) & 7280 (144) \\
\hline
   & T & 1.01 (0.10) & 1.80 (0.08) & 14.21 (0.10) & 7380 (208) \\
37 & M & 1.09 (0.07) & 1.93 (0.06) & 14.20 (0.10) & 7310 (135) \\
   & S & 1.06 (0.06) & 1.84 (0.06) & 14.25 (0.05) & 7335 (122) \\
\hline
   & T & 0.96 (0.07) & 1.73 (0.06) & 14.14 (0.07) & 7167 (102) \\
50 & M & 1.02 (0.06) & 1.71 (0.06) & 14.21 (0.05) & 7290 (91) \\
   & S & 0.97 (0.11) & 1.66 (0.12) & 14.19 (0.07) & 7358 (189) \\
\hline
   & T & 0.97 (0.07) & 1.70 (0.07) & 14.26 (0.06) & 7437 (96) \\
60 & M & 0.96 (0.07) & 1.66 (0.08) & 14.28 (0.04) & 7325 (91) \\
   & S & 0.94 (0.06) & 1.63 (0.05) & 14.23 (0.05) & 7283 (73) \\
\hline

\end{tabular}
\end{table*}

\subsubsection{Protein fluctuations}

\begin{figure}[h!]
\includegraphics[width=7.2cm,clip=true]{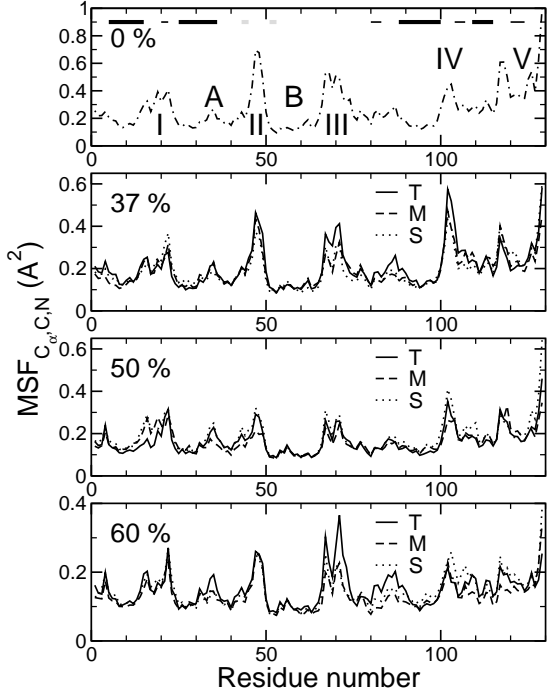}
\caption{\label{u2}
Residue-averaged mean square fluctuations (MSFs) of atoms from the backbone of lysozyme ($C_{\alpha}$,$C$,$N$) in the 
lysozyme/pure water solution and in the different ternary solutions (the overall translational and rotational
motions of the protein have been removed using a least-square fitting procedure). 
The thick and thin black horizontal lines that appears in the top of the figure (in the 0 wt \% plot) 
denote $\alpha$- and 3-10 $\alpha$ helices, respectively, whereas the thick grey lines indicate the $\beta$-sheet.}
\end{figure}

The influence of sugars on the internal motions of lysozyme has been investigated by the computation of atomic 
mean-square fluctuations (MSFs). Similarly to Maragliano~\textit{et al.}~\cite{Maragliano2004}, MSFs were calculated 
by averaging over 250-ps time blocks to limit the effect of conformational changes during simulations. The 
figure~\ref{u2} presents the MSFs of lysozyme main-chain atoms (C$_{\alpha}$, C, and N) for the different studied 
systems, averaged by residue after the removal of the overall translational and rotational
motions of the protein using a least-square fitting procedure.
These MSFs are in qualitative agreement with the experimental temperature factors of 
lysozyme in its tetragonal form~\cite{Young1994}. The high diversity of fluctuations reflects the heterogeneity of local 
environments experienced by lysozyme residues. This is consistent with the three kinds of residues identified by 
Lumry~\cite{Lumry1995}, based on temperature factors, and also with the \textit{knot}/\textit{matrix} classification in 
the description of Gregory~\textit{et al.}~\cite{Gregory1997} from exchange rate distributions~: residues with low and 
intermediate B-factors are identified to \textit{knot} and \textit{matrix} residues, respectively, whereas residues 
with the highest B-factors are located on the surface of lysozyme. In figure~\ref{u2}, five main parts of lysozyme 
(labelled from I to V) have rather high MSFs. These residues correspond to loops, 3-10 
$\alpha$-helices, and some residues of the $\beta$-sheet. They are all located at the protein surface, and are 
therefore likely to form HBs with the solvent. Two zones of the lysozyme residue sequence (labelled A and B) 
have lower MSFs than the average. They correspond to an $\alpha$-helix and a loop located in the core of lysozyme. 
These residues are not extensively accessible to the solvent (see section~\ref{SASA}) and their motions are sterically 
constrained by the presence of other protein residues. The addition of sugars significantly reduces lysozyme 
fluctuations, but their distributions remain qualitatively the same for all sugar concentrations. The MSFs do not reveal 
clear differences among the three sugars, even if maltose tends to globally reduce slightly more protein 
fluctuations than trehalose and sucrose. This result appears in fair agreement with the figure 1a 
of ref.~\cite{Hedoux2006_2}, where the heat capacity at constant pressure $C_{p}$ of the ternary lysozyme/maltose/water 
solution at 40 wt \% is found lower than the $C_{p}$ of the corresponding sucrose and trehalose solutions and may thus 
indicate that lysozyme is less flexible in presence of maltose than in presence of sucrose or trehalose. 
The strong decrease of lysozyme atomic fluctuations with addition of sugars suggests that its dynamics is slaved to 
that of the solvent (see section~\ref{dynamics}) \textit{i.e.} governed by the solvent viscosity. This
observation at 300 K implies that sugars may be able to shift the denaturation temperature $T_{m}$ of proteins towards
higher temperatures, by reducing the amplitudes of local motions that might lead to denaturation. This is actually what
has been observed on lysozyme and other proteins~\cite{Bhat2003,Hedoux2006_2}.

\newpage

\subsubsection{Fractional solvent accessibilies $f_{sa}$}

\label{SASA}

\begin{figure}[!htbp]
\includegraphics[width=7.2cm,clip=true]{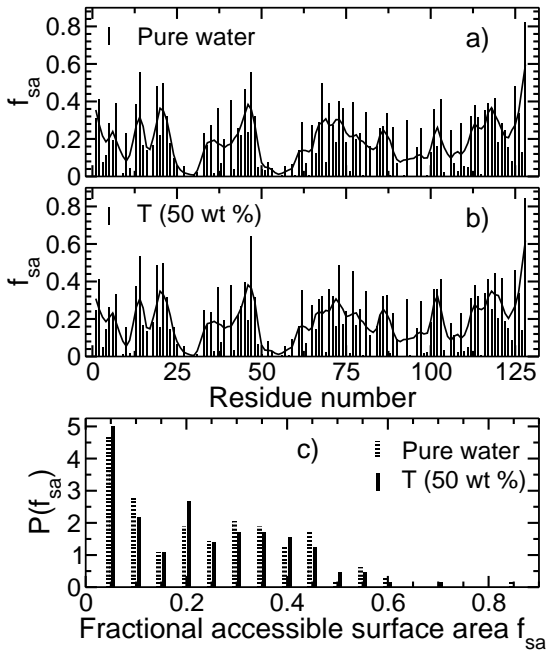}
\caption{\label{FSAres}
Fractional solvent accessible surface areas $f_{sa}$ as a function of the residue number sequence for the lysozyme/pure water (a) 
and for the 50 wt \% trehalose (b) solutions (smoothed curves computed with the Loess algorithm~\cite{Cleveland1979} 
serve as guides to the eye). The related probability distribution functions are displayed in (c).}
\end{figure}

The fluctuations of lysozyme residues clearly reveal distinct local environments, which may arise from differences in 
local packing and interactions with the solvent. Indeed, residues from the core should be more densely packed and less 
accessible to the solvent than residues from the surface, whose motions are also less sterically constrained. The 
time-averaged fractional solvent accessibilities~\cite{Smolin2004} $f_{sa}$ of lysozyme residues have been calculated 
to know how they may interact with the solvent. The DSSP program has been used for this purpose~\cite{Kabsch1983}. 
$f_{sa}$ is equal to the ratio $A_{protein}/A_{free}$, where $A_{protein}$ is the SASA of a particular residue in the 
presence of the other surrounding residues of lysozyme, while $A_{free}$ is the related surface area of the 
\textit{free} residue, \textit{i.e.} without the presence of other amino acids. Figure~\ref{FSAres} shows the $f_{sa}$ 
of the different lysozyme residues in pure water and in the 50 wt \% trehalose solution (the distributions for the 
remaining solutions are similar and are thus not shown). These distributions 
are close to each other and further evidence that the presence of sugars do not modify 
significantly the native conformation of lysozyme. This is well in line with the results from Raman scattering measurements 
in the amide I region of lysozyme (1550-1750 cm$^{-1}$, see figure 5 of ref.~\cite{Hedoux2006_2}), which 
showed that spectra in presence or in absence of sugars at 40 wt \% are very similar to each other. 
This is expected since sugars are well-known protein stabilizers. 
Indeed, according to the preferential hydration hypothesis~\cite{Timasheff2002}, sugars, and more generally osmolytes,
destabilize much more the denatured state of proteins than their native state, and therefore lead to their 
stabilization against the denaturation process.
A more important point is that the distributions mimic those of MSFs (see figure~\ref{u2}), and thus confirm that residues 
with low solvent accessibilities fluctuate much less than residues exposed to the solvent. Figure~\ref{FSAres}c  
presents the distributions probabilities of $f_{sa}$. These distributions show a bimodal behavior~: (i) a first 
relatively sharp contribution for $f_{sa}$ lower than about 0.15, and (ii) a second much broader band, centered at 
$\sim$ 0.3. These two contributions have been attributed to lysozyme \textit{core} and \textit{surface} 
residues, respectively. The sharper contribution from core residues may indicate that the environments they experience 
is much homogeneous than that of surface residues. This again shows that surface residues may exhibit larger motions 
than core residues, because sterical constraints imposed by other protein residues are lower. This result also suggests 
that the exposition of residues to the solvent may have important consequences on their dynamical properties 
(see section~\ref{dynamics}). A detailed analysis of the solvent properties is therefore needed to 
understand how sugars may preserve proteins against denaturation.

\vspace{5.0cm}

\subsection{Protein-solvent interactions.}

\subsubsection{Protein-solvent HBs}

\label{HB_crit}

The way sugars interact with biomolecules, and in particular with globular proteins and membrane 
phospholipid bilayers, has been largely debated over the last two decades~\cite{Crowe1998}. According to the 
\textit{water replacement} hypothesis, sugars form HBs with biomolecules by substituting to their 
hydration water and may then preserve their native structure. 
In the present study, the HBs formed at the lysozyme-solvent interface have been
analyzed. Two molecules were considered to be H-bonded if the donor-acceptor distance $d_{DA}$ is 
less than 3.4 \AA \ and if the D-H$\cdot\cdot\cdot$A angle is larger than 120 deg.~\cite{Liu1997}. 
The table~\ref{HBs} summarizes the time-averaged numbers of HBs formed between lysozyme and solvent
molecules. In the investigated concentration range, sugars form between about 10 and 
20 \% of the total number of protein-solvent HBs, as indicated in table~\ref{HBs}. 
This is much fewer than the number of HBs they could form, assuming that water 
and sugar molecules were equally able to form HBs with lysozyme. Indeed, assuming that water is 
able to form 4 HBs (2 as donor and 2 as acceptor) and that each hydroxyl group of the disaccharides 
may optimally form 3 HBs (1 as donor and 2 as acceptor), the proportion of sugar-lysozyme HBs among 
solvent-lysozyme HBs should be about 15.4, 23.8 and 32.0 \% for the 37, 50, and 60 wt \% solutions, 
when sugar-lysozyme HBs via ring and glycosidic oxygen atoms of sugars are neglected. Therefore, 
sugars are preferentially excluded from the surface of lysozyme, as suggested by the preferential 
hydration hypothesis~\cite{Timasheff1995,Timasheff2002}. In addition, the total number of 
protein-solvent HBs tends to slightly decrease when sugar concentration increases. This suggests 
that sugars are not able to substitute perfectly to water molecules. They indeed cannot access as 
easily as water molecules to polar groups of lysozyme because of sterical hindrance and 
topological constraints.

The number of disaccharide-protein HBs and the number of water molecules shared between lysozyme 
and disaccharides increase with sugar concentration, since sugar-protein interactions are more 
likely. Slight differences are observed among the three sugars. First, sucrose molecules seem to 
form a lower number of HBs with lysozyme in the 50 wt \% solution. This difference may partly 
arise from the fact that the HB network of sucrose is less developed than those of trehalose and
maltose, which have almost percolated at this concentration (see figure~\ref{Cluster}d). It is also 
likely to occur from a lack of configurational sampling of lysozyme-sugar interactions. 
Indeed, sucrose form more comparable numbers of HBs with lysozyme at lower or higher concentrations. 
Consequently, care must be taken when comparing the results for the three sugars. Moreover, it does not seem 
that trehalose forms more HBs with lysozyme than sucrose and maltose at the studied concentrations, 
although this may be the case in the completely dehydrated case. This result would be consistent 
with the simulation results of Lins~\textit{et al.}~\cite{Lins2004} which showed that trehalose 
clusters at the protein surface and does not expel the water molecules closest to the protein
surface. It would therefore confirm the \textit{preferential hydration} hypothesis proposed by 
Timasheff~\textit{et al.}~\cite{Xie1997,Timasheff2002}, which suggests that trehalose is more excluded - in the 
relatively diluted solutions - than other osmolytes, and would thus increase the thermodynamical 
stabilization of the proteins compact native state relative to their extended denaturated 
state~\cite{Wang1999}. The larger hydration number of trehalose (see figure~\ref{Cluster}a) would then 
explain why it does not interact more than do maltose and sucrose with lysozyme in our simulations. 
These results also appear in line with those of a recent simulation study~\cite{Pereira2006}, which 
shows that maltose forms a larger number of HBs with a phospholipid membrane than trehalose, at a 
2 m concentration, at both 325 and 475 K.

Finally, the water replacement hypothesis does not seem confirmed by these results, since 
sugar-protein HBs are not able to substitute perfectly to water-protein HBs, 
probably because of topological constraints and excluded volume effects. Nevertheless, 
sugar-lysozyme HBs relax slower than water-lysozyme HBs as a consequence of the lower mobility 
of sugars, so that a fewer
number of sugar-protein HBs may actually increase the stabilization of proteins. 

\begin{table*}[htbp]
\centering
\caption{\label {HBs}
Mean numbers of water-lysozyme and disaccharide-lysozyme HBs, $n_{HB}$(W-L) and $n_{HB}$(D-L), respectively, 
proportion of disaccharide-lysozyme HBs among solvent-lysozyme HBs 
(\% HB (D-L) = $n_{HB}$(D-L)/($n_{HB}$(D-L)+$n_{HB}$(W-L))) and number $n_{W}$(L-W-D) of water molecules 
which are simultaneously hydrogen-bonded to lysozyme and disaccharides. Standard deviations from mean values 
are given in parentheses.
}
\vspace* {1.0cm}
\begin{tabular}{cccccc}
\hline
$\phi$ (wt \%) &  & n$_{HB}$(W-L) & n$_{HB}$(D-L) & \% HB (D-L) & n$_{W}$(L-W-D) \\
\hline
0 & & 329.0 (5.1) & - &  -  & - \\
\hline
   & T & 291.8 (6.0) & 33.9 (3.2) & 10.4 (1.0) & 46.2 (4.4) \\
37 & M & 285.7 (5.6) & 45.4 (4.7) & 13.7 (1.4) & 54.2 (4.0) \\
   & S & 277.1 (5.6) & 35.5 (3.3) & 11.4 (1.1) & 49.4 (3.6) \\
\hline
   & T & 261.4 (4.5) & 50.9 (3.3) & 16.3 (0.9) & 62.4 (3.8) \\
50 & M & 265.1 (4.2) & 54.9 (4.1) & 17.2 (1.2) & 62.2 (6.7) \\
   & S & 281.5 (5.8) & 34.1 (3.7) & 10.8 (1.1) & 57.3 (4.3) \\
\hline
   & T & 259.7 (5.6) & 57.0 (4.2) & 18.0 (1.3) & 73.0 (3.2) \\
60 & M & 247.2 (6.1) & 62.1 (4.1) & 20.1 (1.4) & 74.4 (5.1) \\
   & S & 243.6 (6.6) & 64.0 (4.1) & 20.8 (1.4) & 65.7 (3.9) \\
\hline

\end{tabular}
\end{table*}

\newpage

\subsubsection{Preferential hydration}

\begin{figure}[h]
\includegraphics[width=7.2cm,clip=true]{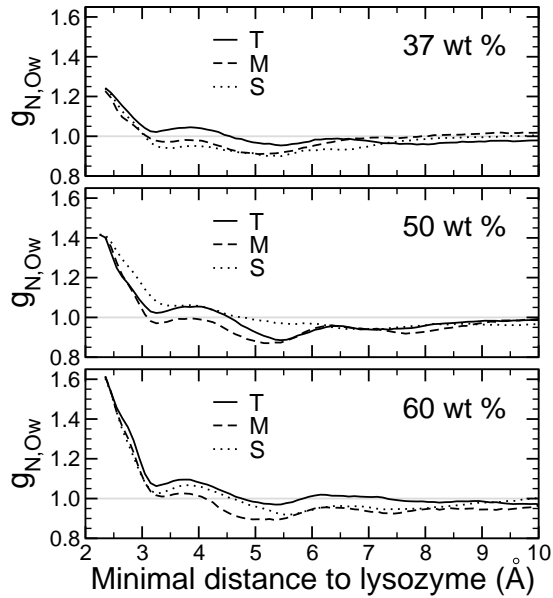}
\caption{\label{gr}
Normalized fraction of water oxygen atoms $g_{N,Ow}$=$n_{OW}/(n_{OW}+n_{OS})$/($N_{OW}/(N_{OW}+N_{OS})$) 
as a function of the minimal distance to any heavy atom from lysozyme. 
$n_{OW}$ and $n_{OS}$ denote the local numbers of water and sugar hydroxyl oxygen atoms, respectively.
$N_{OW}$ and $N_{OS}$ correspond to the total numbers of water and sugar hydroxyl oxygen atoms found in the simulation
box, respectively.}
\end{figure}

Solvent molecules do not form HBs with apolar groups of lysozyme. Therefore, the number of 
protein-solvent HBs does not describe exhaustively protein-solvent interactions. We have thus 
characterized the relative local distribution of water molecules around lysozyme in a similar way to 
Cottone~\textit{et al.}~\cite{Cottone2005}. We have indeed computed the time-averaged normalized ratio 
$g_{N,Ow}$=$n_{OW}/(n_{OW}+n_{OS})$(r)/($N_{OW}/(N_{OW}+N_{OS})$), where $n_{OW}$ and $n_{OS}$ are the local numbers 
of water oxygen atoms and sugar hydroxyl group oxygens, respectively, located at a minimum distance $r$ from 
any heavy atom of lysozyme, and $N_{OW}$ and $N_{OS}$ denote the total numbers of water oxygen atoms and 
sugar hydroxyl group oxygen atoms, respectively, in the simulation box. This ratio is greater than one in 
the close proximity of the protein surface if a given sugar is preferentially excluded from the protein 
surface (in other words, if the protein is preferentially hydrated). Conversely, the sugar preferentially 
interacts with the protein if this ratio is lower than one. This ratio is represented in the figure~\ref{gr} 
for the different ternary studied systems. Lysozyme clearly appears more and more preferentially hydrated 
when increasing sugar concentration. At distances larger than $\sim$ 5 \AA \, there is a slight water 
depletion which results from the presence of sugars. This preferential hydration probably arise from excluded 
volume effects and is consistent with the relatively low proportion of sugar-lysozyme HBs relative to 
water-lysozyme HBs (see table~\ref{HBs}). This may confirm that sugars are good protein stabilizers in 
the framework of the preferential hydration hypothesis~\cite{Timasheff2002}. 
If we exclude the sucrose solution at 50 wt \%, for which there 
may be shortcomings, 
we note that trehalose preferentially hydrates lysozyme slightly more than do maltose and sucrose.
This seems to be especially true for the hydration of apolar groups of lysozyme, which is apparent in the small
peak located around 3.7 \AA \ . Trehalose hydroxyl oxygens were indeed systematically more excluded from these
groups than those of maltose and sucrose (data not shown). This may arise from the larger hydration number 
of trehalose (see figure ~\ref{Cluster}a), which would prevent it from remaining close to apolar groups of lysozyme.
Cottone~\textit{et al.}~\cite{Cottone2005} have also shown that sucrose and trehalose at a concentration
of 89 wt \% preferentially hydrate carboxy-myoglobin (MbCO), but sucrose to a larger extent than trehalose. 
It is possible that the effect of trehalose is somewhat different at this much more elevated 
concentration, as exemplified by the differences between the water replacement~\cite{Crowe1998} and the preferential 
hydration~\cite{Timasheff2002} hypotheses.

\newpage

\subsection{Solvent structure.}

\begin{figure}[h]
\includegraphics[width=7.2cm,clip=true]{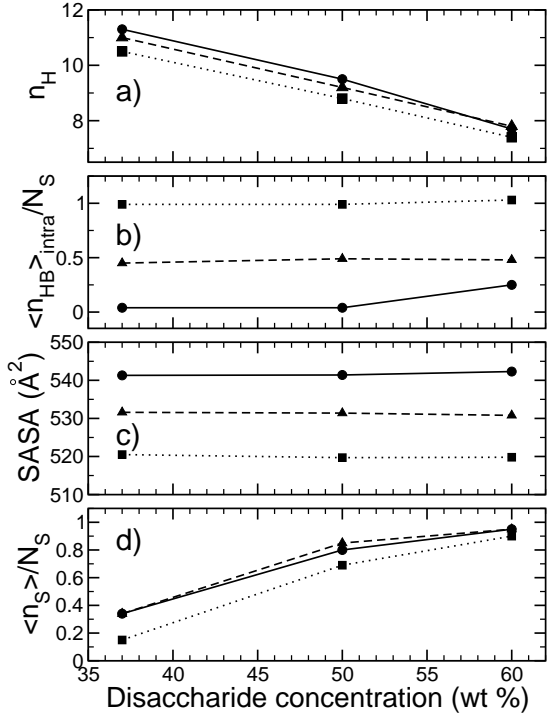}
\caption{\label{Cluster}
Hydration number $n_{H}$ (a), normalized mean number of intramolecular HBs $<n_{HB}>_{intra}/N_{S}$ (b),
solvent accessible surface area (SASA) (c) and normalized mean sugar cluster size
$<n_{S}>/N_{S}$ (d) of trehalose (solid line), maltose (dashed line), and sucrose (dotted line) as a function 
of their concentration.}
\end{figure}

An important issue of the bioprotection phenomenon involves the role of sugar-water interactions. 
Indeed, the stabilizing effect of solutes is sometimes thougth to depend directly on their effect 
on the HB network of water, although this is controversed~\cite{Batchelor2004}. Therefore, the larger 
hydration number of trehalose compared to maltose and sucrose~\cite{Branca2001,Furuki2002,Lerbret2005_2} 
may explain its enhanced stabilizing effect. In the present solutions, the interaction of sugars with water 
molecules is rather analogous to that of binary water/sugar solutions~\cite{Lerbret2005_2}. 
First, the figure~\ref{Cluster}a reveals that the hydration numbers of sugars are lower than in our previous 
simulation study of binary sugar/water mixtures~\cite{Lerbret2005_2}, even if we exclude the sugars hydrogen 
bonded to lysozyme (data not shown).
The analysis of additional simulations of dilute binary sugar/water solutions at a concentration of 4 wt \% 
(1 sugar with 512 water molecules) using the CSFF carbohydrate force field~\cite{Kuttel2002} suggests that 
the observed differences between binary and ternary solutions do not arise from the different force fields, 
Ha~\textit{et al.}~\cite{Ha1988} and CSFF~\cite{Kuttel2002}, used for representing sugars in
ref.~\cite{Lerbret2005_2} and in the present study, respectively. It could thus rather be attributed to the 
preferential exclusion of sugars, which raises their concentrations within their corresponding accessible volumes, 
thus reducing their hydration numbers. Trehalose is found to have a slightly larger hydration 
number than maltose and sucrose, with the exception of 
the 60 wt \% solution, where it is found slightly less hydrated than maltose, as shown in the figure~\ref{Cluster}a.
It is possible that the hydration behavior of sugars in ternary solutions cannot be perfectly extrapolated from
that in binary aqueous solutions, because there should exist for sugars at high concentrations a compromise between 
having a large hydration number and being highly preferentially excluded from the protein surface.
Secondly, the numbers of sugar intramolecular HBs, 
shown in figure~\ref{Cluster}b, are also in qualitative agreement with our previous study~: trehalose is found to 
form a lower number of intramolecular HBs than does sucrose, maltose being intermediate, whatever the concentration
considered. This would explain why the hydration number of trehalose is slightly larger than those of maltose and
sucrose, since hydroxyl groups involved in these HBs no longer remain available for interacting with water molecules.
These numbers definitely depend on the force field used and are systematically found lower in the present
study. This is consistent with the larger radii of gyration of sugars with the CSFF force field~\cite{Kuttel2002}
(3.49 \AA, 3.52 \AA, 3.41 \AA \ for T, M, S, respectively) than with the
Ha~\textit{et al.} force field~\cite{Ha1988} (3.40 \AA, 3.46 \AA, 3.30 \AA \ for T, M, S, respectively) in 
the dilute binary solutions (4 wt \%). 
This result suggests
that sugar conformations are sligthly more extended and that their rings are slightly further from each other 
with the CSFF force field. 
Moreover, the SASAs of sugars computed according to the Lee and Richards method~\cite{Lee1971} with a probe 
radius of 1.4 \AA \ are shown in figure~\ref{Cluster}c. The SASA of trehalose is found larger than that of 
sucrose, that of maltose being intermediate. This property
may explain why trehalose interacts more with water than do maltose and sucrose~\cite{Branca2001,Furuki2002}. 
Besides, the presence of a furanose ring in 
sucrose (see figure~\ref{sugars}) probably explain why several of its properties differ from those of 
maltose and trehalose. 

Finally, the percolation of the HB network of sugars was suggested to play a major role in the bioprotection 
phenomenon, and differences between the three sugars have been observed~\cite{Lerbret2005_2}. This is particularly 
true for sugar concentrations above 50 wt \%, where sugar-sugar interactions become important. 
The normalized mean sugar cluster size $<n_{S}>/N_{S}$ has been computed in the same way as in our previous 
work~\cite{Lerbret2005_2} to know if the percolation of the sugar HB network occurs in the same concentration range 
in presence of the protein. These sizes are presented in the figure~\ref{Cluster}d. A transition clearly appears in 
the [37-50] wt \% concentration interval and the HB network of sugars has almost completely percolated at 60 wt \%, 
in good agreement with our previous results~\cite{Lerbret2005_2}. Sucrose on one hand and maltose and trehalose on 
the other hand show quite different behaviors. Indeed, the percolation transition in sucrose systems is shifted to 
higher concentrations compared to maltose and trehalose ones. In other words, a higher concentration of sucrose is 
needed to reach a given cluster size $<n_{S}>$. This is well in line with the higher number of intramolecular HBs in 
sucrose (see figure~\ref{Cluster}b), as well as its slightly lower accessible surface area due 
to the presence of a furanose ring (see figure~\ref{Cluster}c). 
This behavior may then explain why the dynamics of lysozyme is less modified by the 
presence of sucrose than by the presence of maltose or trehalose, as shown in the section~\ref{dynamics}. 
Furthermore, trehalose and maltose, which are 
topologically closer to each other than to sucrose (see figure~\ref{sugars}), behave in a similar way. 
Nonetheless, it seems that the HB 
network of maltose tends to percolate at a lower sugar concentration than that of trehalose, as observed 
previously~\cite{Lerbret2005_2}. 
This may originate both from the less symmetric conformation of maltose and from its lower affinity for water 
molecules. In our previous study, maltose formed systematically larger clusters than trehalose, whereas this not 
always remains true in the present study. This difference may originate from the presence of the protein, with which 
maltose seems to interact more strongly than trehalose does and thus less with 
other maltose molecules. Interestingly, the mean sugar 
cluster sizes $<n_{S}>/N_{S}$ seem to be consistent with the solubility of the three disaccharides. Indeed, the 
experimental solubility at room temperature of sucrose is significantly larger than that of trehalose, whereas the 
solubility of maltose is slightly lower than that of trehalose (see fig. 4 of ref~\cite{Lammert1997}). Recently, 
Giuffrida~\textit{et al.}~\cite{Giuffrida2006} have suggested that this enhanced tendency of maltose molecules 
to form clusters could be related to their larger dipole momentum. Indeed, from ab initio calculations at the 
BLYP/6-31G** level, they obtained values of 5.2, 2.5 and 1.5 Debyes for the crystallographic structures of maltose, 
sucrose, and trehalose, respectively.

\section{Dynamical properties.}
\label{dynamics}

It is well known that proteins exhibit extremely diverse motions (from local atomic to subunit motions) 
which occur on time scales that span many orders of magnitude~\cite{McCammon1984,Hill2005}.
The broad distribution of the MSFs and of the $f_{sa}$ of lysozyme residues actually 
suggest that their structural relaxations are very heterogeneous, and thus that lysozyme dynamics is complex.
A recent study of hydrated lysozyme ($h$ = 0.43) indeed shows that the distribution of the effective rotational
correlation times of methyl groups is very broad~\cite{Roh2006}.
Furthermore, the effect of sugars on the MSFs of lysozyme suggest that they slow down its
diffusive motions. This would imply that they make lysozyme less sensitive to high temperatures by preventing 
large-scale motions that might lead to denaturation. In order to get a deeper insight 
into the influence of sugars on the relaxational behavior of lysozyme, we have computed 
incoherent intermediate scattering functions $S_{inc}(Q,t)$ for each residue of lysozyme. $S_{inc}(Q,t)$ is defined as :

\begin{equation}
S_{inc}(Q,t) = \bigr < \sum_{\alpha} b_{\alpha,inc}^{2} e^{i.Q.[r_{\alpha}(t)-r_{\alpha}(0)]} \bigl > \\
\label{Sqt}
\end{equation}
where $ b_{\alpha,inc}$ and $r$ denote the incoherent scattering length and the vector position of a 
given atom $\alpha$, respectively, and $Q$ is the wavevector - the brackets mean averaging over every 
time origin of simulations. These functions represent the time Fourier transform of dynamic structure
factors $S_{inc}(Q,\omega)$, obtained in neutron scattering experiments. We have chosen to
probe the dynamics of lysozyme at a wavevector of 2.29 \AA$^{-1}$, which corresponds to the position 
of the first peak in the static structure factor $S_{O-O}$ of pure water and which was used in our 
study of binary sugar/water solutions~\cite{Bordat2004}. Cha\-rac\-teris\-tic relaxation times $\tau$ may then be defined 
as the decay times from 1 to 1/$e$ of the different $S_{inc}(Q,t)$. 

\begin{figure}[h]
\includegraphics[width=7.2cm,clip=true]{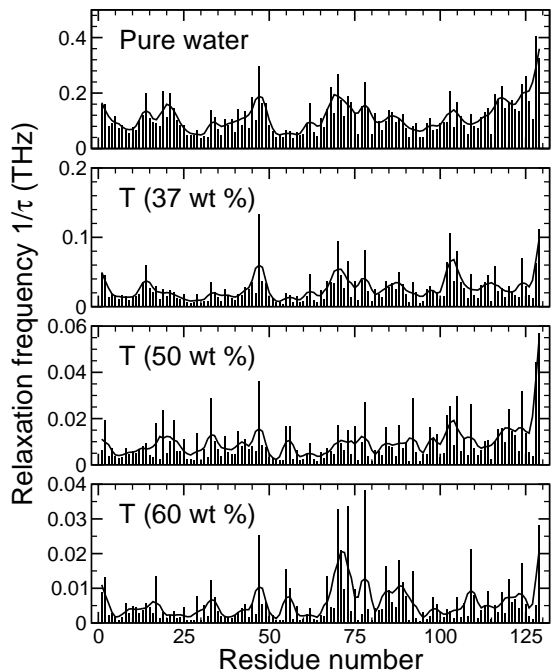}
\caption{\label{Taures}
Relaxation frequencies 1/$\tau$ of the residues of lysozyme in pure water and in the different 
trehalose solutions (smoothed curves computed with the Loess algorithm~\cite{Cleveland1979}
serve as guides to the eye).
}
\end{figure}

The figure~\ref{Taures} shows the relaxation frequencies 1/$\tau$ of residues of
lysozyme in pure water and in the different trehalose solutions. These frequencies clearly appear diverse 
and, similarly to MSFs, they seem to follow the fractional solvent accessibilies $f_{sa}$ of residues of
lysozyme (see figure~\ref{FSAres}), since residues which are more exposed to the solvent relax faster than those which
are buried in the protein \textit{core}. 

The addition of sugars at the studied concentrations induces a strong reduction of the relaxation frequency 
of lysozyme residues. This slowing-down of the dynamics of lysozyme evidences that the higher the solvent 
viscosity, the lower the diffusive (or $\alpha$-like) motions of lysozyme, in the studied concentration 
range. This would confirm the basic assumption that the structural or $\alpha$ relaxation of the solvent 
- which is related to its viscosity - determines the influence of the solvent on the dynamics of 
proteins~\cite{Walser2001}.

\begin{figure}[h]
\includegraphics[width=7.2cm,clip=true]{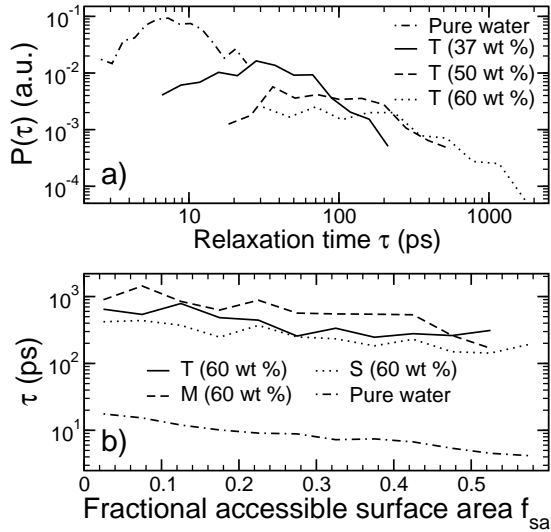}
\caption{\label{Taures2}
(a) Probability distribution functions of the relaxation times $\tau$ of lysozyme residues in the pure 
water/lysozyme solution and in the different trehalose solutions.
(b) Averaged relaxation times of lysozyme residues $\tau$ as a function of their fractional solvent 
accessibilities $f_{sa}$ for the pure water solution and for the 60 wt \% trehalose, 
maltose and sucrose solutions.}
\end{figure}

The dynamical slowing-down induced by sugars is more visible in the figure~\ref{Taures2}a, which shows the 
probability distribution functions of the relaxation times
$\tau$ of lysozyme residues in the pure water and in trehalose solutions (the distribution functions for
maltose and sucrose solutions exhibit comparable behaviors and are not shown for clarity reasons).
A significant shift towards larger relaxation times as well as a broadening of distributions are observed when
the sugar concentration increases. This may reflect the broadening of the energy barriers experienced by lysozyme
residues resulting from an increased dynamical coupling between the solvent and the protein.
Furthermore, the correlation between $f_{sa}$ and relaxation times $\tau$ remains relatively well 
established 
at the highest studied concentration (60 wt \%), as shown in figure~\ref{Taures2}b, where is represented the 
mean relaxation time of lysozyme residues as a function of their fractional accessible surface area $f_{sa}$, for
$f_{sa}$ up to of about 0.6 (given the low number of residues with a $f_{sa}$ larger than 0.6 as shown 
in figure~\ref{FSAres}, the related statistical errors are very large). 
This figure shows that sugars influence both the dynamics of 
\textit{core} and \textit{surface} residues. It seems therefore possible to modulate the internal dynamics 
of proteins by changing the solvent viscosity with the addition of sugars. 

\begin{figure}[h]
\includegraphics[width=7.2cm,clip=true]{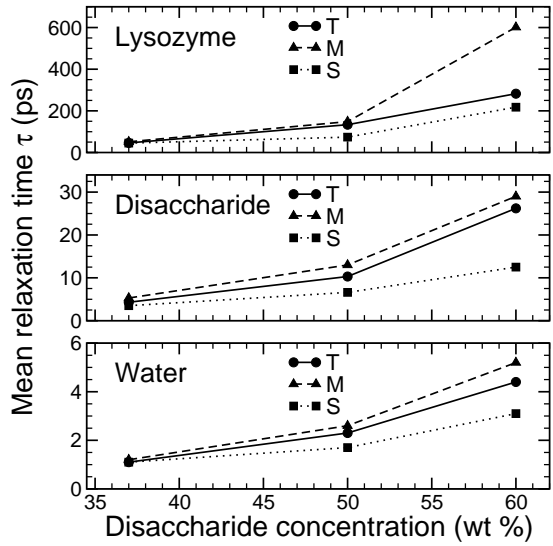}
\caption{\label{taudecomp}
Mean relaxation times $\tau$ of lysozyme (top), disaccharides (middle) and water (bottom) as a function of the 
disaccharide concentration.}
\end{figure}

Figure~\ref{Taures2}b also shows that the relative influence of sugars on the dynamics of lysozyme
does not change too much over the entire $f_{sa}$ range. Therefore, a straigthforward comparison between 
sugars is possible by considering the mean relaxation time
of the whole lysozyme. The mean relaxation times of sugar and water molecules have also been computed to
interpret the effect of sugars on lysozyme and solvent dynamics.
The figure~\ref{taudecomp} shows that the dependences on the disaccharide concentration of the mean relaxation times 
of lysozyme, sugars and water molecules exhibit a similar slope change at a concentration close to 50 w \%. This 
may be interpreted when considering the percolations of the sugar HB networks, which occur at a concentration of 
about 40-50 wt \% in both 
the sugar binary~\cite{Lerbret2005_2} and ternary solutions (see figure~\ref{Cluster}d). At these relatively high 
concentrations, the hydration layers of sugars do not surround completely disaccharides because of sugar-sugar HBs,
which strongly influence their dynamics. These dynamical changes have been observed experimentally by
Rampp~\textit{et al.}~\cite{Rampp2000} for a series of carbohydrates - including sucrose and trehalose - at
concentrations above 50 wt \%. They also appear in the concentration dependences of the diffusion coefficient of
sucrose and trehalose in aqueous solutions obtained by Ekdawi-Sever~\textit{et al.} by NMR measurements~\cite{Feick2003}. 
Sugar-sugar interactions are not large enough at 37 wt \% for detecting significant differences 
between sugar solutions. But at 50 wt \%, the HB network of maltose and trehalose is more developed than that of sucrose, 
so that each species relaxes more slowly in the maltose and the trehalose solutions than in the sucrose one. 
Finally, the HB network of sugars has almost percolated in the 60 wt \% solutions. Their dynamics are then
dominated by the relaxation of the HB network of sugars and differences among sugars emerge. Sucrose clearly relaxes 
faster than trehalose and maltose, due to its lower ability to form clusters.
This is well in line with experimental studies, which show that the dynamics of sucrose is faster than that of 
maltose and trehalose~\cite{Branca2001}. Moreover, the relaxation times of maltose seem to be somewhat longer 
than those of trehalose. This appears consistent with the lower heat capacity at constant pressure $C_{p}$ of the 
ternary lysozyme/maltose/water solution at 40 wt \% in comparison with the corresponding sucrose and trehalose 
solutions~\cite{Hedoux2006_2}, which suggests that the ternary maltose solution is more viscous. This result also 
seems to confirm the \textit{peculiar behavior} of maltose observed by 
Giuffrida~\textit{et al.}~\cite{Giuffrida2006}, where the matrix dynamics was found the most reduced in the 
\textit{dry} maltose sample. Experimental measurements suggest that binary trehalose/water solutions 
are sligthly more viscous than maltose ones~\cite{Branca2001,Matsuoka2002}. Therefore, the present results may 
underline the influence of the protein on the dynamics of the solvent (see \textit{e.g.} ref.~\cite{Dirama2006} and 
ref.~\cite{Marchi2002}), which may arise from (i) the roughness of the protein surface, (ii) a decrease of the dimensionality 
of the solvent diffusion at the protein-solvent interface and (iii) strong solvent-protein interactions with polar and 
charged groups of the protein. Given that the conformation of lysozyme is similar in trehalose and maltose solutions, 
the stronger interaction of maltose with lysozyme (see table~\ref{HBs}) may imply that its dynamics is more slowed down 
than that of trehalose.

Furthermore, a direct comparison between the expected biopreservative efficiencies of the three studied carbohydrates 
in the light of the present results is not straightforward since maltose is a reducing sugar, unlike trehalose and sucrose. 
It may thus react with lysine and arginine residues of lysozyme and other proteins to form carbohydrate adducts, 
especially at high temperatures, via a complex browning pathway called Maillard reaction~\cite{Wang1999} (which cannot 
occur in MD simulations). 
This may explain why maltose is rarely used experimentally for protein stabilization, in contrast to trehalose 
and sucrose, which are natural bioprotectants. Maltose was actually found to induce the smallest shift of 
the denaturation temperature $T_{m}$ of lysozyme at a concentration of 40 wt \%, as measured in modulated 
differential scanning calorimetry and Raman scattering investigations~\cite{Hedoux2006_2}. 

In addition, figure~\ref{taudecomp} reveals that the protein, sugar and
water dynamics are slower in the trehalose solutions than in the sucrose solutions for concentrations above 40-50 wt \%.
These results are in agreement with the tighter coupling of carboxy-myoglobin (MbCO) with trehalose-water matrices
than with sucrose-water matrices observed experimentally by Giuffrida~\textit{et al.} by means of Fourier transform 
Infrared (FTIR) measurements~\cite{Giuffrida2004}. This enhanced slowing-down presumably 
arises from the 
larger sugar-sugar interactions in the trehalose solutions at these concentrations, as
seen in figure~\ref{Cluster}d. Moreover, the larger hydration number of trehalose compared to sucrose allows a more
important dynamical coupling with water molecules. Therefore, trehalose would have a better preservation efficiency
because of the quasi-absence of intramolecular HBs in its solution conformation 
(see figure~\ref{Cluster}b), which allows it to interact more strongly 
with both water and sugar molecules than sucrose does. On the other hand, this would explain its higher solubility 
than (true) maltose~\cite{Lammert1997}. This suggests that the greater \textit{homogeneity}~\cite{Lerbret2005_2} 
of intermolecular interactions in the trehalose solutions (water-trehalose and trehalose-trehalose) might lead to 
their better preservation 
efficiency. This may explain why trehalose was found to induce the largest shift of $T_{m}$ and 
to increase the most the temperature interval of the denaturation process $\Delta T$ of lysozyme~\cite{Hedoux2006_2}.
The shift of $T_{m}$ was indeed found to be of 9.4$\pm$0.3 K, 4.2$\pm$0.3 K and 2.9$\pm$0.3 K in the trehalose, sucrose 
and maltose solutions, respectively, by Raman scattering. Similarly, the temperature domain $\Delta T$ of the denaturation 
process was extended by 0.9$\pm$0.2 K, 0.4$\pm$0.2 K and 0.5$\pm$0.2 K in the trehalose, sucrose and maltose solutions, 
respectively (see table 2 of ref.~\cite{Hedoux2006_2}).

Finally, it must be pointed out that the slowing-down of lysozyme dynamics follows the slowing-down 
of the solvent induced by the percolation of the
sugar HB network, but not necessarily the number of lysozyme-sugar HBs. Indeed, sucrose forms at 60 wt \% a larger number 
of HBs with lysozyme than trehalose, but lysozyme relaxes faster in the sucrose solution. This would imply that the
slowing-down of the solvent plays a more important role on the dynamics of lysozyme than specific sugar-lysozyme
HB interactions in the studied concentration range. If we only compare trehalose and sucrose, this suggests that
the hypothesis of Green and Angell~\cite{Green1989}, as well 
as the preferential hydration hypothesis~\cite{Timasheff1995,Timasheff2002} might be valid for this
kind of solutions. In presence of sugars, the unfolding process would require a reorganization of the HB network of 
sugars, easier in the sucrose solutions than in the trehalose ones.

\newpage

\section{Conclusions}

This article reports the results from molecular dynamics simulations of lyso\-zyme/di\-saccha\-ride/water solutions
for intermediate concentrations of disaccharides (37-60 wt \% on a protein-free basis). 
This concentration range was chosen because it 
corresponds to the range of concentrations where the HB network of sugars was shown to percolate in our previous
study of disaccharide/water solutions~\cite{Lerbret2005_2}. Several parameters of lysozyme like its 
root-mean-square deviations (RMSD)
from the crystallographic structure, its radius of gyration and its SASA 
indicate that its conformation in presence of sugars remains very similar to that in pure water. 
Nevertheless, structural fluctuations of lysozyme as seen from mean-square fluctuations (MSFs) are much 
reduced in presence of sugars. This reduction was shown to be quite homogeneous throughout the protein and follows
the fractional solvent accessibilities $f_{sa}$ of lysozyme residues. This points out the influence of the solvent on
the internal dynamics of lysozyme. The analysis of the interactions of sugars with lysozyme via HBs as well as the relative
concentration of water oxygen atoms around lysozyme suggest that lysozyme is preferentially hydrated, well in line 
with other studies~\cite{Lins2004,Cottone2005}. When comparing the three sugars, it seems that trehalose is 
slightly more excluded than maltose and sucrose, although the observed differences may be not significant. This could be 
interpreted by the larger hydration number of trehalose compared to that of maltose and 
sucrose~\cite{Branca2001,Furuki2002}. Furthermore, the number 
of intramolecular HBs of trehalose was found lower than that of maltose and sucrose. 
Interestingly, trehalose was also found to have the largest SASA. This may explain why it interacts more with 
water than maltose and sucrose. In addition, the HB network 
of sugars was shown to percolate in the studied concentration range. Sucrose forms smaller clusters than trehalose 
and maltose, probably because of its larger number of intramolecular HBs. 
Dynamical properties of lysozyme have been analyzed via incoherent intermediate scattering functions $S_{inc}(Q,t)$.
The relaxation times of lysozyme residues mimic both MSFs and $f_{sa}$ distributions and a broadening of the distributions 
of relaxation times of lysozyme residues is observed when increasing sugar concentration. This could reveal an enhanced dynamical
coupling between solvent and lysozyme. Moreover, a similar increase of lysozyme, sugar and water relaxation times 
is observed when increasing the sugar concentration from 37 to 60 wt \% and may arise 
from the percolation of the HB network of sugars, as suggested from our previous results on binary sugar/water 
solutions~\cite{Lerbret2005_2}. Since sucrose forms smaller
clusters than trehalose and maltose, it appears to slow down the dynamics of lysozyme less strongly at concentrations
above $\sim$ 40-50 wt \%. This appears consistent with experimental results, which sugggest a looser 
protein-solvent coupling in sucrose-water matrices than in trehalose-water matrices 
of various water contents~\cite{Giuffrida2004}. 
In contrast, maltose slows down more efficiently the dynamics of lysozyme because it forms 
larger clusters and interacts more strongly with lysozyme than trehalose. This seems consistent with the lower
heat capacity at constant pressure $C_{p}$ of lysozyme/maltose/water solutions at 40 wt \% compared to that 
of trehalose and sucrose solutions~\cite{Hedoux2006_2} and with the \textit{peculiar behavior} of maltose 
observed experimentally by Giuffrida~\textit{et al.}~\cite{Giuffrida2006}.

Our simulations only cover a limited concentration range of disaccharides at ambient temperature (300 K), so that
lysozyme is not submitted to temperature and/or desiccation stresses. Furthermore, the length
of simulations is too short to account for phenonema which occur on much longer time scales in real systems, such as
vitrification, ice formation or sugar crystalline phase transitions. Nevertheless, our results suggest that the 
dynamical slowing-down of lysozyme induced by the presence of sugars may stabilize globular proteins against
thermal denaturation. In addition, our results indicate that lysozyme remains preferentially hydrated in the
studied concentration range, since the substitution of water-lysozyme HBs by sugar-lysozyme HBs is rather limited.

The comparison of the expected biopreservative behaviors of the three studied disaccharides on lysozyme 
is not straightforward since maltose is a reducing sugar and may thus alter the structure of lysozyme 
when chemically reacting with
its lysine and arginine residues via the Maillard reaction. Consequently, the
comparison of sucrose and trehalose, which are naturally synthesized osmolytes, is more relevant in the context 
of preserving globular proteins.
The results of this study indicate that trehalose has a stronger influence on the dynamics of water and lysozyme than
sucrose and may therefore preserve more efficiently the native structure of lysozyme at higher temperatures,
particularly at concentrations above $\sim$ 40-50 wt \%. These differences appear to primarily stem from their 
different topologies (see figure~\ref{sugars}), which 
cause the formation of intramolecular HBs in sucrose and - almost - none in trehalose.
The main consequences are (i) a lower number of water molecules with which sucrose molecules may interact and 
(ii) the formation of sucrose clusters of lower sizes, which would have thus a reduced influence 
on the solvent and lysozyme dynamics. Even though these differences between trehalose and sucrose appear relatively small at
room temperature where lysozyme is in its compact native state, they may increase during the unfolding process of lysozyme 
and then play a major role. \\

{\bf Acknowledgments}
                                                                                                                                                             
The authors wish to acknowledge the use of the facilities of the IDRIS (Orsay, France), the CINES (Montpellier, France) and
the CRI (Villeneuve d'Ascq, France) where calculations were carried out. This work was supported by the INTERREG III (FEDER) 
program (Nord-Pas de Calais/Kent).

\bibliographystyle{jpc}

\end{document}